\documentclass[12pt]{article}
\usepackage{graphicx}
\usepackage{multirow}
\usepackage{hyperref}
\usepackage{epsfig}
\usepackage{amsmath}

\begin{document}

\title{Nonlinear Effects of Gravity in Cosmology}
\author{Johan Hansson\footnote{c.johan.hansson@ltu.se}, Jaime Dols Duxans \& Martin Svensson \\
 \textit{Division of Physics} \\ \textit{Lule\aa \,University of Technology}
 \\ \textit{SE-971 87 Lule\aa, Sweden}}

\date{}

\maketitle

\begin{abstract}
We consider some nonlinear effects of gravity in cosmology. 
\\
Possible physically interesting consequences include: non-requirement of dark matter and dark energy, asymmetric gravitational matter-creation, emergent homogeneity/isotropy \& asymptotic flatness, resolution of ``cosmic coincidence" $\Omega_m \sim \Omega_{\Lambda}$, effective cutoff of gravitational interaction at the scale of cosmic voids.
\end{abstract}
\section{Introduction}
The standard Friedmann-Robertson-Walker(FRW)-model of cosmology is overly simplified as it assumes perfect, and eternal, spatial homogeneity and isotropy (``cosmological principle") - something simply not observed in the physical universe. The largest known structure so far is around 3000 Mpc \cite{Horvath}, comparable to the size of the observable universe and clearly in conflict with the assumed ``cosmological principle" of FRW.

The lure of FRW is that these extreme simplifying assumptions reduce Einstein's equations from a generally intractable set of ten coupled, highly nonlinear, partial differential equations to an analytically solvable set of just two coupled ordinary differential equations (Friedmann's equations), with only one dynamical degree of freedom, the cosmic scalefactor $a(t)$.

However, gravitational energy itself gravitates, so in the real ``lumpy" universe we get ``runaway solutions" where gravity nonlinearly amplifies the effect of matter, the most extreme example being the formation of a black hole which in the end continues to exist solely due to gravitational nonlinearity.
As it long has been observationally known that there are huge spongy structures in the universe; cosmic ``filaments" and ``voids" \cite{Geller}, where the former contain all observable matter, whereas the latter seem essentially empty of such, these nonlinear effects must play a major role for the dynamics and observations of the universe, ever more so as structure formation proceeds.

\begin{figure}
\begin{center}
\scalebox{.4}{\includegraphics {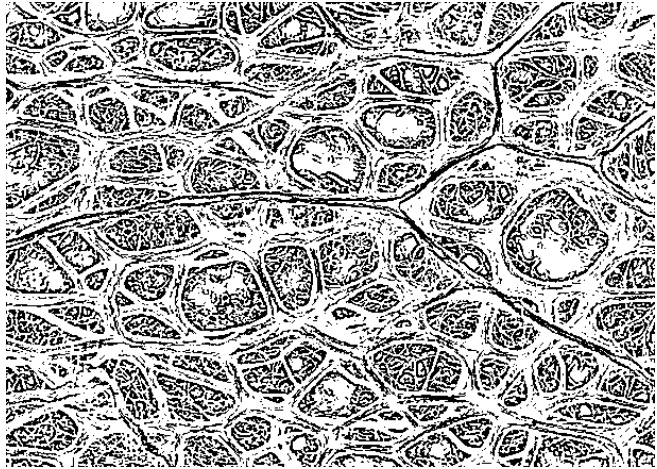}}
\end{center}
\caption{Schematic representation of large-scale spatial structure, with gravitationally bound regions of overdensity surrounding gravitationally unbound, expanding regions of underdensity. Just like the soap in the bubbles in a bubble-bath is distributed in a way to minimize the nonlinear energy, something similar should explain the observed filaments and voids in the cosmos (global energy being ill-defined in general relativity).}
\end{figure}
\section{General Relativity}
The cosmos, according to the purely classical theory general relativity, is an unchanging ``block" of 4D-spacetime, in which all events in the global universe are embedded.
We have for the global scalar curvature invariant (proportional to the Einstein-Hilbert action of general relativity, and related to the Yamabe invariant of differential geometry)
\begin{equation}
\int_{tot} R \sqrt{|g|} \, d^4x = A,
\end{equation}
where $R = R^{\mu}_{\mu} = g^{\mu \nu}R_{\mu \nu}$ is the scalar curvature, and $g$ the determinant of the metric.
For our present classically causal (observable) universe since the big bang the 4-volume is finite. Its 3D spatial hypersurfaces are non-simply connected, with evolving stringy regions of overdensity, surrounding bubbly regions of underdensity, Fig. 1. For the gravitationally bound regions of structure (observationally known to be almost ``fractal-like"/scale-invariant and hierarchical),
\begin{equation}
\int_b R \sqrt{|g|} \, d^4x = B,
\end{equation}

and for gravitationally unbound regions, using (1), we must have
\begin{equation}
\int_u R \sqrt{|g|} \, d^4x = U = A-B.
\end{equation}
From Einstein's equations
\begin{equation}
R_{\mu \nu} - \frac{1}{2} g_{\mu \nu} R = - \frac{8 \pi G}{c^4} T_{\mu \nu},
\end{equation}
where $T_{\mu \nu}$ encodes all non-gravitational energy-momentum and stress, the 4D scalar curvature always fulfills
\begin{equation}
R^{\mu}_{\mu} - \frac{1}{2} g_{\mu \nu}g^{\mu \nu} R = - \frac{8 \pi G}{c^4} T^{\mu}_{\mu},
\end{equation}
\textit{i.e.}
\begin{equation}
R = \frac{8 \pi G}{c^4} T.
\end{equation}
And as we never have perfect matter-less vacuum in the real universe, only in idealized models, the scalar curvature is generally non-zero, and positive definite in the sign convention we use. So far the discussion has been exact and generally covariant/invariant.

Now approximating, for a perfect fluid in comoving coordinates in homogeneous and isotropic space
\begin{equation}
T = \rho(t) c^2 - 3p(t).
\end{equation}
Hence, for an eternally expanding global FRW-universe $T$, and thus $R$, goes to zero as $t \rightarrow \infty$.
For pure radiation $p = \rho c^2 /3$, giving $R_{rad} = 0$, so radiation does not contribute to the mean spacetime curvature.
(Generally, the energy-momentum tensor of a pure electromagnetic radiation field is traceless so $T_{rad} = 0$ always.)
For a present matter-dominated epoch $p \simeq 0$ (``dust") and $R_{matter} = 8 \pi G \rho(t) / c^2$.
If a cosmological constant, $\Lambda$, is introduced in (4) via a term $\Lambda g_{\mu \nu}$,
\begin{equation}
R = \frac{8 \pi G}{c^4} T - 4\Lambda,
\end{equation}
then
$R \rightarrow - 4\Lambda = const.$ as $t \rightarrow \infty$, making (1) and the Einstein-Hilbert action ill-defined as the 4-volume goes to infinity; a purely mathematical
reason to avoid $\Lambda$ quite apart from the physical reasons: $\sim 10^{120}$ discrepancy between the theoretically expected value of $\Lambda$ and the observationally inferred one, no known physical fields can generate $\Lambda$, two vastly different $\Lambda$'s at inflation and now, etc.

As all particles are hyper-relativistic in the early universe, they can then be treated as radiation and thus $T_{early} \rightarrow T_{rad} = 0 \Rightarrow R_{early} \rightarrow 0 \Rightarrow$ (Einstein-Hilbert Action)$_{early} \rightarrow 0$, explaining why the universe started out smooth on the average. Furthermore, as $T \rightarrow 0$ also as $t \rightarrow \infty$, (1) and the global action is non-divergent, provided that $\Lambda =0$.

As a quantum treatment is necessary when the action over a characteristic 4-volume is less than or $\sim \hbar$, we see that a universe containing pure radiation (as in the earliest epochs, and the latest ones if the universe expands forever and black holes evaporate) really would have to be treated quantum gravitationally; only in the intermediate epochs does the classical treatment apply. This also admits the possibility of small (quantum) inhomogeneities/fluctuations in the early universe, whereas the classical treatment would be smooth, precluding ``seeds" from which gravitational structures later could grow.We thus see that quantum considerations is a necessity for cosmology, not a luxury.

To study dynamics, \textit{i.e.} evolution, we must make a 3+1-split of spacetime, which at the same time destroys the globally invariant character of the 4D spacetime ``block". However, cosmological structure formation introduces a natural ``arrow of time" all by itself - the standard big bang model being isotropic in its spatial part only, not in time.

The scalar curvature in a perfectly homogeneous and isotropic (FRW) universe is just a function of the cosmic scalefactor $a=a(t)$
\begin{equation}
R = R^0_0 + R^1_1 + R^2_2 + R^3_3 = \frac{6}{c^2} (\frac{\ddot{a}}{a} + \frac{\dot{a}^2}{a^2} + \frac{k c^2}{a^2}),
\end{equation}
and
\begin{equation}
R^0_0  = \frac{3}{c^2} \frac{\ddot{a}}{a},
\end{equation}
\begin{equation}
R^1_1 = R^2_2 = R^3_3 = \frac{1}{c^2} (\frac{\ddot{a}}{a} + \frac{2 \dot{a}^2}{a^2} + \frac{2 k c^2}{a^2}),
\end{equation}
where $k = {+1, 0, -1}$ characterizes the spatial curvature of the FRW-universe.

More generally, for gravitationally nonlinearly bound spatial regions, Fig. 1, we have
\begin{equation}
\int_b R^{(3)} \sqrt{|g^{(3)}|} \, d^3x = C,
\end{equation}
where $R^{(3)}$ is the intrinsic scalar curvature for the spatial hypersurface at a given epoch, and $g^{(3)}$ the determinant of its induced three-metric. We must, by necessity, be situated in a region where gravitationally bound structures have formed, and so do all objects we observe, which is not characteristic of the global physical universe.
For the gravitationally unbound spatial regions in Fig. 1 we have
\begin{equation}
\int_u R^{(3)} \sqrt{|g^{(3)}|} \, d^3x = D,
\end{equation}
where in a globally spatially flat universe
\begin{equation}
C + D = \int_{tot, flat} R^{(3)} \sqrt{|g^{(3)}|} \, d^3x.
\end{equation}
As structure formation proceeds and the universe expands the cosmic ``voids" increasingly contribute the bulk of the physical volume\footnote{As a non-expanding two-dimensional analogy, a flat sheet of metal if punched with small, sharp indentations ``compensates" by curving the other way in between. (Even the original ``flat" sheet is, if magnified sufficiently, seen to consist of negatively and positively curved regions.) Similarly, the surface of the earth globally has positive curvature, even though locally there are regions of negative curvature such as saddle-shaped mountain passes. If we do not assume global flatness cosmologically, we will have a different right-hand side in Eq.(14). However, in the universe, global \textit{spatial} flatness translates to zero energy in the newtonian (weak-field) setting, making it ``cost nothing" to create such a universe, and making it possible for it to exist indefinitely; $\Delta E \Delta t \sim \hbar$, $\Delta E \rightarrow 0 \Rightarrow \Delta t \rightarrow \infty$, where asymptotically the newtonian limit becomes exact. In that sense a flat universe would not be merely a possibility, requiring fine-tuning of initial conditions, but a necessity.}, cosmologically mimicking the effects of a mysterious, but fictitious, dark energy as the relative expansion rate between voids and filaments increases. As gravity is increasingly concentrated along filaments it is ``diluted" in the voids, Eqs. (12)-(14).
We empirically know that the ``lumpiness" of the observed universe has increased with time, and that the effect thus monotonously increases, at least until very late epochs far removed from the present. From the reasoning above it then becomes natural that we should observe an (apparent) ``acceleration" only in an epoch with appreciable void/filament structure formation ($z \sim 1$ and below). In the much simplified FRW standard model of cosmology with non-zero cosmological constant $\Lambda$, on the other hand, it is just inexplicable why the ``cosmic coincidence" $\Omega_{matter} \sim \Omega_{\Lambda}$ should occur \textit{now}, as $\Omega_{matter} \gg \Omega_{\Lambda}$ as $t \rightarrow 0$ and $\Omega_{matter} \ll \Omega_{\Lambda}$ as $t \rightarrow \infty$. Also, to obtain the observed large-scale structure in a ``mere" 14 billion years, the standard FRW-model must be ``doped" by huge amounts of dark matter ($\Lambda$CDM-model), much larger than the contribution from known ``normal" matter, as newtonian gravity used in N-body simulations to study supposedly ``small" perturbations in the assumed globally valid background FRW-geometry does \textit{not} generate additional gravity (Poisson's equation is \textit{linear}).
In full, nonlinear general relativity the ``doping" may be automatically provided by gravity itself - as gravity now \textit{does} generate additional gravity \cite{Thorne} it will nonlinearly amplify the gravity of the matter present in bound regions. Also, as the field equations are nonlinear, solutions do not superpose, unlike gravitational potentials in the newtonian approximation. The field of two galaxies is \textit{not} the sum of the ``individual" fields, let alone the huge number used in (linearized) N-body simulations. In that sense, structure formation in the universe, both in space and time, may be seen as just one more example of spontaneous self-organization, known to require nonequilibrium and nonlinearity \cite{Prigogine}. One should keep firmly in mind that the dark matter has never been seen, and is completely hypothetical. There is, in fact, not a single shred of independent evidence for either dark matter or dark energy outside of astronomy/cosmology.

\section{Matter Creation}
It is known that the only other strongly nonlinear fundamental interaction with massless force-carrier particles, Quantum Chromodynamics (QCD), supports ``string-like" characteristics of the effective force ($\neq 1/r^2$) due to the nonlinearity of the color interaction. If we assume that the same occurs for gravity, as it too couples nonlinearly to itself, and as a first approximation assume that the effect scales with the intrinsic strength between two protons (hydrogen being the main component of the visible universe) $F_{QCD}/F_{gravity} \sim 10^{38}$, we get for the present characteristic length-scale of Gravity-string-alterations $10^8$ lightyears, which is intriguing, as it is of the same order as the size of cosmic filaments and voids (the effect needing time, \textit{i.e.} space, to become effective). Due to the strongly nonlinear nature, even very small initial deviations may then have large effects, especially over vast regions of space and time.

Gravity, as it self-couples, should thus also get a correction to its ``field-lines", see Fig. 2.
\begin{figure}
\begin{center}
\scalebox{.6}{\includegraphics {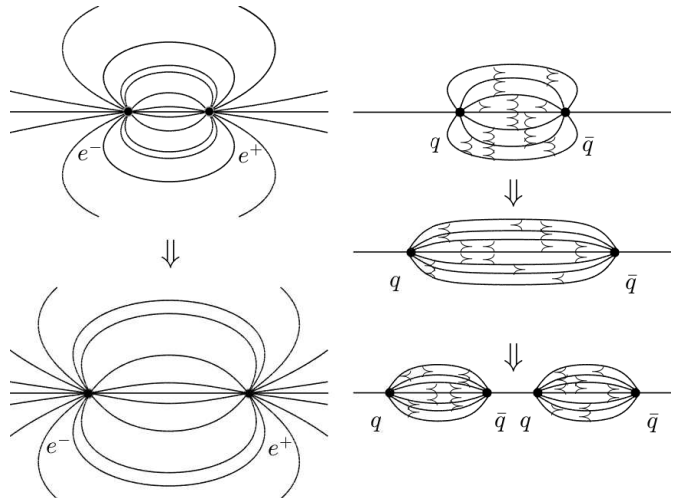}}
\end{center}
\caption{Schematic representation of effect without (QED: left) and with (QCD: right) nonlinear self-interaction. Gravity should behave in a way similar to QCD, as it too is nonlinearly self-interacting. However, at completely different scale. Above a threshold distance the effective force is $\sim$ constant, the potential rising as $\sim r$, until it becomes favorable to create matter, which gives an effective cut-off of the interaction. Heuristically, one can thus immediately appreciate why matter should lie along filaments, surrounding voids, Fig. 1. This should qualitatively be true also in full general relativity, as gravity itself gravitates, but can no longer be described by a simple (single) potential.}
\end{figure}
This could then create matter, akin to Hoyle's old steady-state model with continuous creation \cite{Hoyle}, just like ``hadronization" in QCD: when the flux-tube is stretched it eventually becomes energetically favorable to create new matter particles non-adiabatically out of the field energy. The scale of hadronization in QCD is $\sim$ 10 fm, and as the work needed to create proton-pairs is $F_{QCD} \cdot l_{QCD} \simeq F_{gravity} \cdot l_{gravity} = 2 m_p c^2$, in cosmology matter creation should take place at a scale of roughly $10^8$ lightyears. Gravity then gets an effective cut-off at this scale, just like QCD exponentially disappears outside nuclei (even though gluons are massless), giving an immediate reason as to why the universe should seem statistically homogeneous and isotropic at scales $\gg 10^8$ lightyears.

As general relativity is globally neither CP nor CPT invariant, the CPT-theorem being valid only for Lorentz-covariance \cite{Pauli}, strict particle-antiparticle symmetry need not apply, meaning that CP can be broken, even if T-reversal symmetry is still obeyed. (However, it is the T-symmetry that conserves energy, and this too is broken globally in general relativity.) So CP is broken globally, which means that a global asymmetry between particles and antiparticles is theoretically allowed, as also is observed. It is an empirical fact that the weaker the interaction, the fewer conservation laws does it obey. So this aspect of gravity, the by far weakest of all known interactions, need not be entirely surprising. ``Gravitation conserves everything or nothing depending on how you look at it" \cite{Feynman}. More concretely, this dissipative effect should enhance distribution of matter along the field lines, \textit{i.e.} between the (initially tiny) anisotropic matter concentrations, making them more and more pronounced, qualitatively explaining why matter is observed to increasingly lie along filamentary ``rope-like" structures. At the same time, as gravity effectively concentrates along the filaments it decreases in the voids, making them relatively speed up. For visualization, the familiar 3D ``raising dough"-analogy can still be utilized, but with embedded ``raisins" (galaxies) replaced by a ``flexible web" (filaments) that expands and thins as the voids expand, see Fig. 1. The 2D ``inflating balloon"-analogy can also be used, but with ``coins" (galaxies) glued to the surface replaced by ``rubber bands" crisscrossing the balloon surface. Regions ``in between" (voids) then relatively expand increasingly faster than the ``flexible web"/``rubber bands" (filaments), giving an apparent relative acceleration. As always, these graphic analogies are not exact, and should not be perceived as such, but only as rough guides for the mind.

If matter creation does occur gravitationally, we get a simple and natural explanation of the ``flatness problem", \textit{i.e.} why the universe is perceived to be flat regardless of initial conditions ($\neq$ globally recollapsing). Excess kinetic energy of the matter automatically generates new matter until there is an exact balance between gravitational and kinetic energy, the matter thus automatically approaching critical density globally as the expansion proceeds, the expansion asymptotically going towards zero, just like for a \textit{flat} FRW-universe as $t \rightarrow \infty$.

In QCD, hadronization globally conserves energy and momentum of matter by construction, as in particle physics the, inherently nonlocal, gravitational effect can be (and is) completely neglected, as spacetime is effectively flat to extremely high accuracy in particle physics experiments. This cannot be fulfilled in general relativity, and especially not in cosmology, as: i) global energy is not really defined, ii) the generally covariant four-divergence
$T^{\mu \nu}_{; \nu} = 0$ is not a conservation law but describes the local \textit{exchange} of energy and momentum between (local) matter and (nonlocal) gravitation \cite{Weinberg}, the energy and momentum of matter not being conserved in the presence of dynamic spacetime curvature but changing in response to it.

The equivalence principle itself precludes a local definition of gravitational energy, but if we introduce the gravitational energy-momentum pseudo-tensor $t_{\mu \nu}$ (not generally covariant), then
\begin{equation}
\tilde{T}^{\mu \nu} = T^{\mu \nu} + t^{\mu \nu},
\end{equation}
locally does give a (merely Lorentz-covariant) conservation law
\begin{equation}
\tilde{T}^{\mu \nu},_{\nu} = 0,
\end{equation}
where the comma denotes normal partial (not covariant) differentiation. So, there seems to be no fundamental obstacle in \textit{principle} to gravitational matter creation.

\section{Emergent Homogeneity and Isotropy}
As one expects severe gravitational/metric ``turbulence" \cite{Thorne} in the early universe (as opposed to the comparatively ``laminar" approximate Hubble flow of today) one would expect thorough mixing leading to a high degree of \textit{statistical} homogeneity and isotropy. This would be ``turbulence" in spacetime itself, so naive causal constraints (global FRW-horizons) do not apply, and is automatically embedded in general relativity needing no \textit{ad hoc} mechanism like inflation. If gravitational ``turbulence" is nearly scale-invariant, like turbulence in fluids, it could generate the approximately scale-invariant fluctuations seemingly needed for structure formation in the universe.

As the earliest epoch furthermore must be described by some sort of quantum gravity it should exhibit the same kind of nonlocality that ordinary quantum mechanics is known to do through tests \cite{BellTest} of Bell's theorem \cite{Bell} from 1972 until the present \cite{Aspect}, making classical causal horizons in that epoch lose their meaning and observed global near-isotropy (\textit{e.g.} microwave background) natural. This is based on a \textit{known} property of laboratory-tested physics, and does not rely on speculative (and untested) additional hypotheses about what happens $\leq 10^{-32}$ s after time zero in the very early universe. Also, a pure quantum state of the universe would have zero entropy, beginning to increase only as ``observations" (collapse of the wavefunction, \textit{i.e.} quantum $\rightarrow$ classical transition) occur, connecting that to the cosmological arrow of time, the 2nd law of thermodynamics and structure formation. If the asymptotic energy content of the universe ($t \rightarrow \infty $) furthermore is zero (the universe arising from a quantum fluctuation) it implies global flatness$^1$, and a quantum universe created in its groundstate would be isotropic.\footnote{Making inflation superfluous, which is inconsistent anyway as it is based on classical (hypothetical) fields, assuming definite values at each point in spacetime, contributing to the energy-momentum tensor in an era where quantum effects should reign supreme.}

\section{Dark Matter?}
The dark matter normally inferred in galaxy clusters, \textit{e.g.} through the gravitational lens effect, as a broadly distributed hump devoid of any visible matter, superposed on ``spiky" individual galaxies, may in part or totality be the nonlocal energy ($\neq T_{\mu \nu}$) of the gravitational field itself. It would by definition be ``dark", \textit{i.e.} invisible, as it has no other interactions but gravitational. Furthermore, already well below $10^8$ lightyears, gravity should get corrections to $r^{-2}$ due to the contracting of field-lines, Fig. 2. This would mean that in a spiral galaxy the rotational (disk) plane should experience a stronger effective gravity, making deviations from Keplerian motion natural for large $r$, \textit{without} any dark matter. Using the analogy between gravity and QCD the effective gravitational force, outside some threshold range (and well before matter production, which effectively cuts off gravity), should be $\sim$ constant, \textit{i.e.} with effective gravitational potential
\begin{equation}
\phi \simeq - \frac{GM}{r} + \alpha M r,
\end{equation}
resulting in
\begin{equation}
v  \simeq \sqrt{\frac{GM}{r} + \alpha M r},
\end{equation}
the first term giving the normal Keplerian dynamics for most astrophysical scales, while the second term dominates for large $r$, giving
\begin{equation}
v  \propto \sqrt{r},
\end{equation}
compatible with observations \cite{Mannheim}.
As $\alpha$ starts to dominate when $\alpha M r \geq GM/r$, \textit{i.e.} for $r \geq \sqrt{G/\alpha}$, this could explain galactic dynamics without dark matter if $\alpha$ is of the order $10^{-51}$ N/kg$^2$. The ratio $G/\alpha$ should in principle be calculable from a theory of quantum gravity, Fig. 2, and may even be scale-dependent.

Furthermore, the recently empirically discovered, unexpected, direct and seemingly universal relation between the ``dark matter" and the normal visible baryonic mass in spiral galaxies \cite{McGaugh} could get an automatic explanation as a nonlinear effect of gravity \textit{itself}, without the need for any dark matter. One should keep in mind that such a relation is not at all natural for dark matter models, and was neither anticipated nor predicted by them.

\section{Conclusion}
The traditional approach to cosmology is: 1. Construct an overly simplified global model (FRW, with only \textit{one} degree of freedom). 2. Deduce its dynamical consequences through general relativity (Friedmann's equations) and believe blindly in the conclusions, conveniently ``forgetting" the severe approximations made. 3. Observe that the consequences do \textit{not} correspond to the real universe, especially in the recent era. 4. Invent make-shift add-on ``solutions" (dark matter, dark energy) to save the model - instead of discarding it and constructing a less simplified model.

We, instead, propose that full nonlinear gravity itself has the potential to explain most cosmological observations and currently perceived ``enigmas", without the introduction of new \textit{ad hoc} components of the universe. The arguments and model-calculations in this article at least do not preclude such a possibility.

\end{document}